\documentclass[aps,prl,onecolumn,superscriptaddress]{revtex4-2}

\usepackage{graphicx}
\usepackage{epsfig}
\usepackage{dcolumn}
\usepackage{bm}
\usepackage{overpic}
\usepackage{color}
\usepackage{multirow}
\usepackage{subfigure}
\usepackage{lineno}
\usepackage{amsmath}
\usepackage{url}
\usepackage[table]{xcolor}
\usepackage{diagbox}
\usepackage{multirow}
\usepackage{ulem}

\usepackage{xcolor}

\usepackage{hyperref}
\hypersetup{
colorlinks=true,
linkcolor=blue,
filecolor=blue,
urlcolor=blue,
citecolor=blue,
pdftitle={LaTeX Example},
pdfpagemode=FullScreen,
}

\begin{document}


\title{A novel approach to determine photon polarization at collider experiments}

\author{Xiao-Rong Lv}
\affiliation{Institute of Modern Physics, Chinese Academy of Sciences, Lanzhou 730000, People's Republic of China}

\author{Yu-Tie Liang}
\email[]{liangyt@impcas.ac.cn}
\affiliation{Institute of Modern Physics, Chinese Academy of Sciences, Lanzhou 730000, People's Republic of China}
\affiliation{University of Chinese Academy of Sciences, Beijing 100049, People's Republic of China}

\author{Boxing Gou}
\email[]{gouboxing@impcas.ac.cn}
\affiliation{Institute of Modern Physics, Chinese Academy of Sciences, Lanzhou 730000, People's Republic of China}
\affiliation{University of Chinese Academy of Sciences, Beijing 100049, People's Republic of China}

\author{Chuang-Xin Lin}
\email[]{linchuangxin@impcas.ac.cn}
\affiliation{Institute of Modern Physics, Chinese Academy of Sciences, Lanzhou 730000, People's Republic of China}
\affiliation{University of Chinese Academy of Sciences, Beijing 100049, People's Republic of China}

\author{Ai-Qiang Guo}
\email[]{guoaq@impcas.ac.cn}
\affiliation{Institute of Modern Physics, Chinese Academy of Sciences, Lanzhou 730000, People's Republic of China}
\affiliation{University of Chinese Academy of Sciences, Beijing 100049, People's Republic of China}

\date{\today}

\begin{abstract}

The polarization of final-state photons is a critical observable for probing the fundamental mechanisms of particle and nuclear interactions, providing insights into spin and parity structure that are inaccessible through cross-section measurements alone. However, this observable remains largely unexplored in collider experiments, as general-purpose spectrometers traditionally lack the capability to measure it. This paper proposes a novel technique to integrate photon polarimeter function into such a spectrometer without compromising the spectrometer's conventional performance. Key factors to enhance the polarimeter capability are investigated. This successful integration represents the first implementation of a photon polarimeter within a general-purpose spectrometer, establishing a valuable benchmark for the existing and future experiments. The ability to concurrently measure spin polarization and four-momentum data opens a new dimension for analysis, promising a more profound understanding of the underlying physics.


\end{abstract}

\maketitle

\section{Introduction}

Photon polarization represents a fundamental property of light that has evolved from an optical phenomenon to an essential tool in modern physics. As a transverse wave, the electromagnetic field of a photon oscillates perpendicular to its direction of propagation, with the polarization state describing the orientation and behavior of these oscillations. Linear polarization occurs when the electric field oscillates in a single plane, while circular polarization results when the electric field vector rotates circularly around the propagation direction. In both nuclear decays and high-energy collisions, the polarization of final-state photons encodes critical information about the underlying interaction mechanisms, revealing details of nuclear structure~\cite{RevModPhys.31.711, DROSTE1999260, Zheng:2013jrx}, fundamental symmetries~\cite{Nat_Comm_CP}, and quantum entanglement~\cite{PhysRev.73.440, Nat_Comm_quantum_entanglement}, etc. This intrinsic property transforms photons from mere energy carriers into precision diagnostic tools for fundamental physics.




The systematic measurement of photon polarization began in astrophysics, driven by the need to decode cosmic magnetic fields and acceleration mechanisms. In the 1940s–50s, optical polarimetry of starlight laid the groundwork for polarization-dependent radiative transfer models~\cite{Chandrasekhar_astro_series, Rybicki}. Breakthroughs followed with the discovery of polarized synchrotron emission from the Crab Nebula~\cite{Mayer1958}, confirming relativistic electron trajectories.
In the 21st century, a numbers of balloon or space missions measured the photon polarization from gamma-ray bursts~\cite{McGlynn, Lowell:2017yti, Yonetoku, POLAR_phy}.
 These historical developments in both instrumentation and methodology have established polarimetry as an indispensable technique across multiple domains of physics and astronomy.


Photon polarization measurement techniques diverge sharply across energy regimes due to the evolving dominance of photon-matter interactions. In the region of keV to MeV, a Compton polarimeter can be used, leveraging the azimuthal asymmetry of scattered photons like GRETINA~\cite{MORSE2022166155, LI20188}. For high-energy photons beyond the GeV scale, pair production polarimeters are typically used. These measure the azimuthal asymmetry of the electron-positron pair production planes~\cite{MALDERA2023168080, Gros:2016dmp}, where the recoil nucleus is undetectable. Alternatively, pair production on an atomic electron can be utilized if the recoil electron is detected. In this case, the azimuthal angle of the recoil electron provides a means to extract the photon polarization~\cite{DUGGER2017115}.


In particle and nuclear physics, multi-GeV linearly polarized photon beams are a powerful probe for studying hadronic structure, as demonstrated by programs at JLab~\cite{ESCOFFIER2005563} and SPring-8~\cite{Muramatsu:2021bpl}. While dedicated photon polarimeters measure beam polarization, determining photon polarization remains challenging. For the often-used pair production polarimeter, the primary obstacle is the stringent requirement to resolve the azimuthal angular separation of $e^+e^-$ from pair production. These pairs emerge at exceptionally small opening angles, approaching detector resolution limits. This challenge is particularly acute for final-state photons with uncontrolled trajectories, unlike the collimated primary beam.




Polarization measurement on the final-state particles unlocks a new observational dimension, complementing traditional data on mass spectra and cross-sections. While general-purpose spectrometers in collider experiments in particle and nuclear physics excel at reconstructing the four-momentum of stable particles, they typically lack the capability to measure spin polarization. Inspired by a recent method for determining nucleon polarization without significant performance trade-offs~\cite{c642-1lzb}, this paper proposes a novel approach and optimization direction for spectrometer to enable photon polarization determination. Throughout the paper, photon polarization refers to linear polarization. We present a comprehensive investigation, including the detector concept, the polarization extraction methodology, cross-section and uncertainty estimation.


\section{General principle of photon polarimeter}

\begin{figure}
\includegraphics[width=0.7\textwidth]{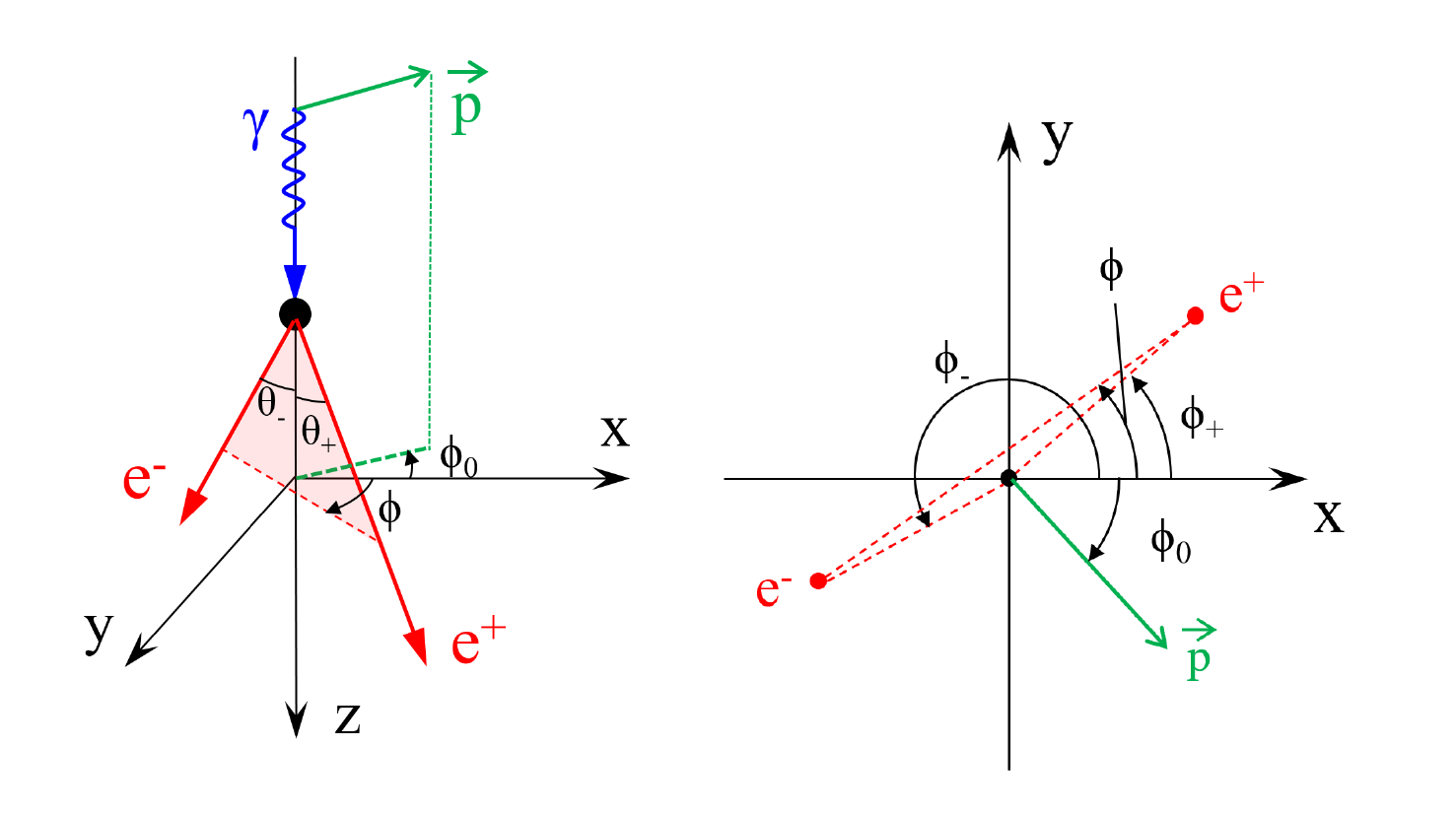}
\caption{\label{fig:gamma_conversion_illustration} {Illustration of gamma conversion to an $e^+e^-$ pair. The directions of positron (electron) momentum are defined by the angles $\theta_{+(-)}$ and $\phi_{+(-)}$. $\phi_0$ is the polarization angle of the incident photon. $\phi$ is the azimuthal angle of the event. 
}}
\end{figure}

For nuclear and particle physics applications, specialized techniques have been developed to measure high-energy photon polarization. At energies around GeV, a typical region of final-state photons in particle physics experiments, photons primarily interact via pair production, as shown in Fig.~\ref{fig:gamma_conversion_illustration}. Pair production polarimeters exploit the correlation between the photon polarization plane and the $e^+e^-$ pair production plane, as first identified in Refs.~\cite{PhysRev.77.722.2, PhysRev.78.623}. The complete quantum electrodynamic analysis of the effects of polarization in pair production was subsequently provided by Olsen and Maximon~\cite{PhysRev.114.887}, yielding practical formulas for the cross-section of pair production that enable the determination of the polarization from the azimuthal asymmetry of the components of the pair.


The differential cross-section for $e^+e^-$ pair production by a linearly polarized photon can be expressed in the form~\cite{Gros:2016dmp, BERNARD201885}

\begin{align}
\frac{\mathrm{d}\sigma}{\mathrm{d}E_+\,\mathrm{d}\Omega_+\,\mathrm{d}\Omega_-} = \frac{-\alpha Z^2 r_0^2}{(2\pi)^2} \frac{|p_+||p_-|m^2 M}{E^3 \sqrt{s}|\vec{q}|^4}(X_u + P X_p)\,
\end{align}

\begin{align}
X_u &= \left(\frac{p_+\sin\theta_+}{E_+-p_+\cos\theta_+}\right)^2(4E_-^2-q^2)\nonumber\\
& +\left(\frac{p_-\sin\theta_-}{E_--p_-\cos\theta_-}\right)^2(4E_+^2-q^2)\nonumber\\
&\left.+\frac{2p_+p_-\sin\theta_+\sin\theta_-\cos(\phi_+-\phi_-)}{(E_--p_-\cos\theta_-)(E_+-p_+\cos\theta_+)}(4E_+E_-+q^2-2E^2)\right.\nonumber\\
&-2E^2\frac{(p_+\sin\theta_+)^2+(p_-\sin\theta_-)^2}{(E_+-p_+\cos\theta_+)(E_--p_-\cos\theta_-)},
\end{align}


\begin{align}
X_p &= \cos2\phi_-(4E_+^2-q^2)\left(\frac{p_-\sin\theta_-}{E_--p_-\cos\theta_-}\right)^2 \nonumber\\
&+\cos2\phi_+(4E_-^2-q^2)\left(\frac{p_+\sin\theta_+}{E_+-p_+\cos\theta_+}\right)^2\nonumber\\
&+2\cos(\phi_++\phi_-)(4E_+E_-+q^2) \nonumber\\
&\times\frac{p_-\sin\theta_-p_+\sin\theta_+}{(E_--p_-\cos\theta_-)(E_+-p_+\cos\theta_+)}.
\end{align}

where $P$ is the linear polarization of the photon; $E$ is the photon energy, while $E_+$ and $E_-$ are the positron and electron energies, respectively; $\mathrm{d} \Omega_+$ and $\mathrm{d} \Omega_-$ are the elements of solid angles for their emission directions; $\alpha$ is the fine structure constant, $r_0$ is the classical electron radius, $q$ is the momentum of the recoil nuclei, and $Z$ is the atomic number of converter material. The term $X_{u}$ represents the cross-section for unpolarized photons, while $X_{p}$ constitutes the polarization-dependent part. 

Integrating over the $E_+$ and polar angle of the electron and positron, one obtains the azimuthal angle dependent cross-section


\begin{equation}
\frac{d\sigma}{d\phi} \propto 1+ \mathcal{A} P\cos[2(\phi-\phi_0)]
\label{eq:pol_XSC}
\end{equation}

Here, $\mathcal{A}$ is the polarization asymmetry of the conversion process which can be calculated. The angle origin, $\phi_0$ determines the orientation of the polarization of the photon flux with respect to a given reference direction. Different definitions of the azimuthal angle can be used, including the azimuthal angle of one detected lepton $\phi_+$ or $\phi_-$, the azimuthal angle of the pair plane $\phi$, or the bisector angle ($\phi_+ + \phi_-$)/2, as illustrated in Fig.~\ref{fig:gamma_conversion_illustration}.


Pair production polarimeters have been established as small-acceptance devices for measuring the linear polarization of photon beams in particle and nuclear physics experiments~\cite{KOBAYASHI1972101, WOJTSEKHOWSKI2003605}. A central challenge for such polarimeters is the precise measurement of the azimuthal angular separation between the electron and positron, which are produced at extremely small opening angles. At 1 GeV, for instance, the peak occurs at approximately 1.5 mrad based on generators in Refs.~\cite{BERNARD201885}.


Historically, various methods have been employed to address this challenge. One approach uses a dipole magnetic field to rapidly separate the pair~\cite{KOBAYASHI1972101}. Another method relies on a long flight distance (on the order of 1 meter) in a magnetic-field-free environment to achieve separation~\cite{WOJTSEKHOWSKI2003605}. However, these techniques—requiring either a dedicated dipole magnet aligned with a fixed photon momentum direction, or a large volume free of magnetic fields—are not suitable for integration into modern general-purpose spectrometers in collider experiments.




\section{Feasibility of general-purpose spectrometer as photon polarimeter}

In particle and nuclear physics experiments at colliders, general-purpose spectrometers are commonly employed, such as Belle-II~\cite{ONUKI202278}, BESIII~\cite{ABLIKIM2010345}, STAR~\cite{Chen:2024aom}, or the future
STCF~\cite{Achasov:2023gey}, EicC~\cite{Anderle:2021wcy}, EIC~\cite{AbdulKhalek:2021gbh}, etc. These spectrometers typically feature a large solenoid magnet and an array of detectors, such as tracking detectors to measure the momentum of charged particles, calorimeters to measure the energy of photons or hadrons, and particle identification detectors like time-of-flight or Cherenkov counters. With a large solid-angle coverage—some approaching a full 4$\pi$ steradians—and the capability to reconstruct the four-momentum of all stable final-state particles, general-purpose spectrometers are widely versatile tools in high-energy physics research.


In these spectrometers, final-state photons can undergo $\gamma$-conversion in detector material, producing $e^+e^-$ pairs. Using the BESIII experiment as an example, photon conversion events occurring at the beam pipe or the inner wall of the drift chamber had been observed, exhibiting a clear conversion vertex signature~\cite{PhysRevD.103.092005}. The trajectories of the charged $e^+e^-$ pair curve in the spectrometer's solenoid magnetic field, allowing their momentum to be reconstructed by the tracking detector. The distinct conversion vertex demonstrates that a general-purpose spectrometer can successfully reconstruct such conversion events. In principle, such a setup could also measure the photon’s polarization if the spectrometer is sufficiently precise to resolve the azimuthal angular distribution of the pair. 
However, the typical angular momentum resolution of general-purpose tracking systems at around several mrad, is comparable to or larger than the typical $e^+e^-$ opening angle, posing significant challenges for resolving polarization-dependent azimuthal separations.

\begin{figure}
\includegraphics[width=0.48\textwidth]{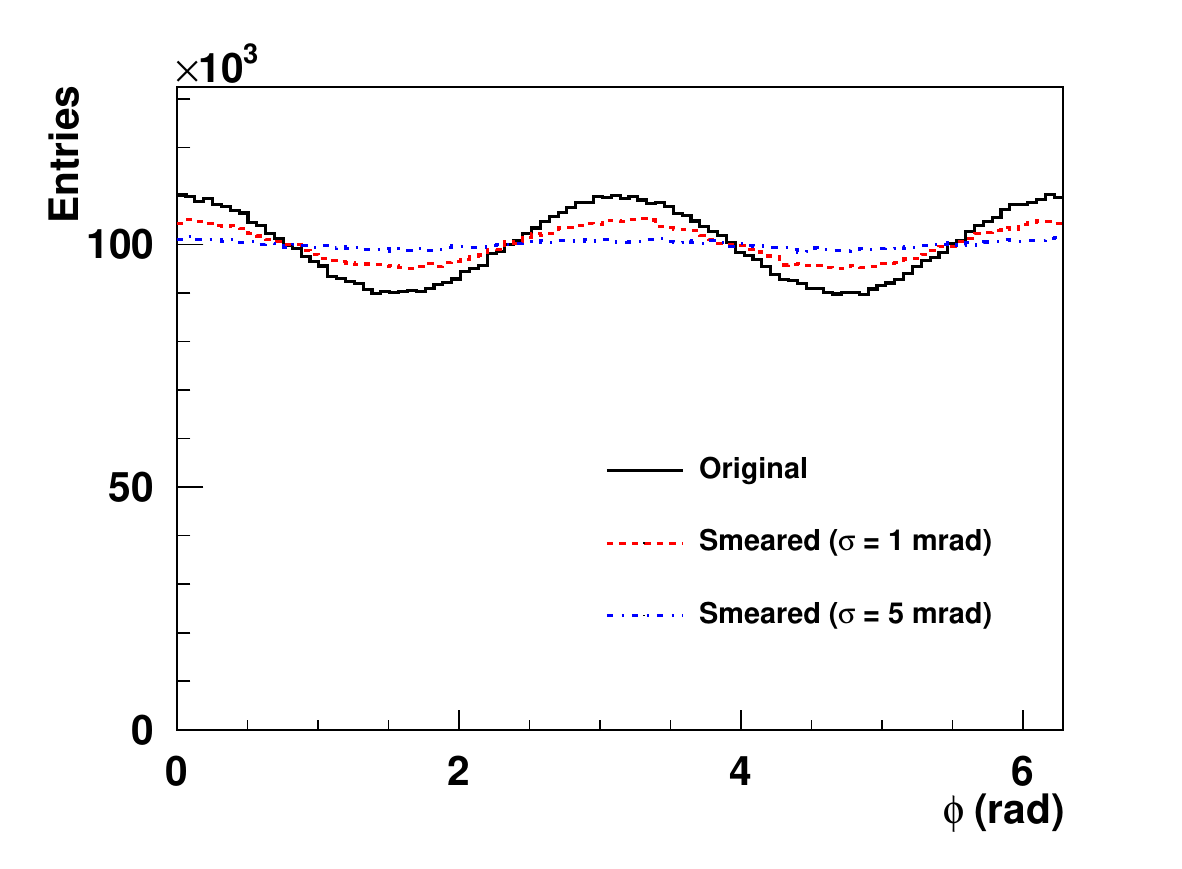}
\caption{\label{fig:smear_effect} {The impact of detector angular resolution. The black line shows the original asymmetry of 0.1 at an opening angle of 1 mrad. The red dashed line indicates the smeared asymmetry with detector angular resolution of 1 mrad, and the blue dot-dashed line is with resolution of 5 mrad.
}}
\end{figure}

A critical issue is that mis-measurement of $\phi$ due to angular resolution dilutes the asymmetry. Figure~\ref{fig:smear_effect} shows a preliminary study of smeared asymmetries at different detector angular resolutions with an opening angle of 1 mrad. The black solid line is the true distribution with an original asymmetry of 0.1. The red dashed line is the smeared distribution with a detector angular resolution of 1 mrad. The blue dot-dashed line is with angular resolution of 5 mrad, in which the decrease of the asymmetry is very pronounced.



This preliminary investigation suggests that the angular momentum resolution of a general-purpose spectrometer is generally insufficient for accurate photon polarization measurements via the pair production process. However, they also reveal that once the angular momentum resolution of the spectrometer is fixed, a predictable relationship emerges between the true polarization and the observed value. This problem could be resolved—and the functionality of a photon polarimeter realized within a general-purpose instrument—by establishing a calibration table to map observed polarization values to their corresponding true values.


\section{Methodology illustration}

\begin{figure}
\includegraphics[width=0.48\textwidth]{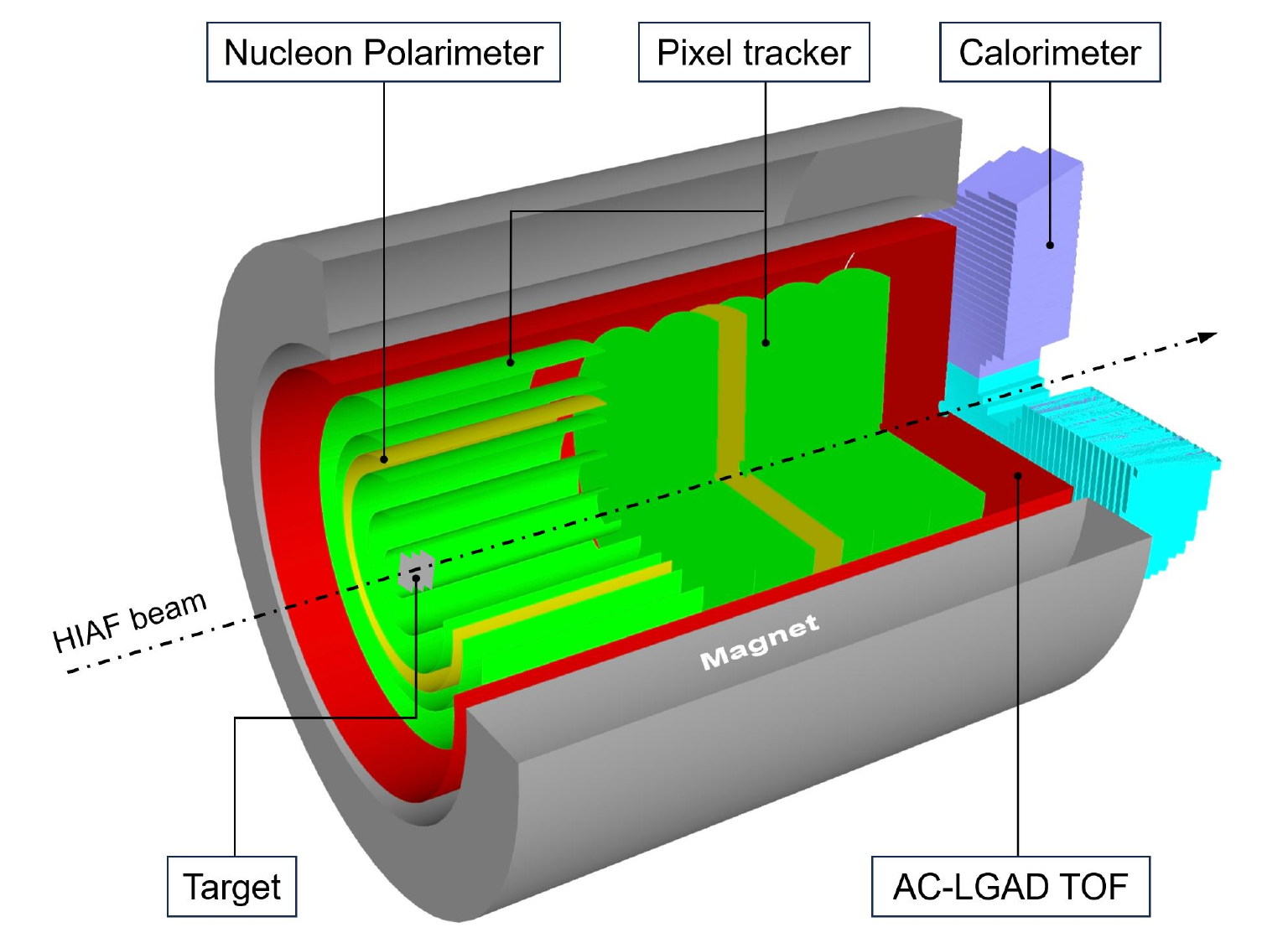}
\includegraphics[width=0.35\textwidth]{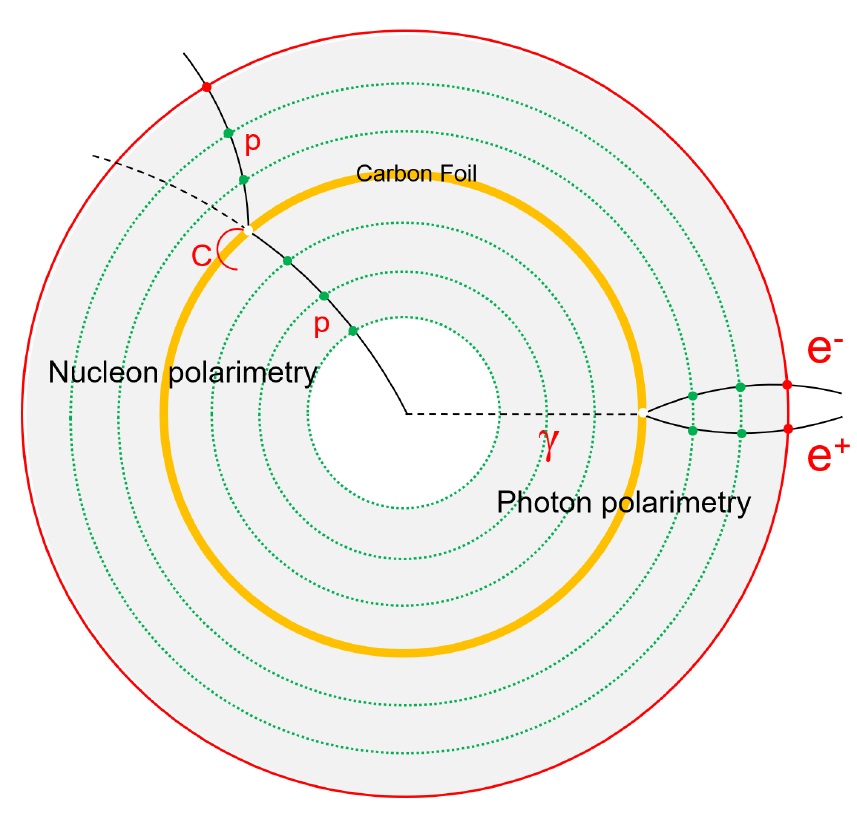}
\caption{\label{fig:HNS_det} {Concept of H-NS detector(Left) and illustration of $\gamma$-conversion to $e^+e^-$ pair on carbon layer in the H-NS (Right). This study focuses on the photons from the primary vertex..
}}
\end{figure}


To evaluate the potential of a general-purpose spectrometer as a photon polarimeter, a case study is performed based on the future Hyperon-Nucleon Spectrometer (H-NS), depicted in the left plot in Fig.~\ref{fig:HNS_det}. The H‑NS features a tracking detector built with modular Monolithic Active Pixel Sensors~\cite{RTurchetta_2006, Herui-NST, DuanFF-NST, Wangshen-NST, Cao-JINST, ZhaoC-JINST, Malong-NST}. Each sensor layer has a pixel size of 30 $\mu$m and contributes a material budget of 0.35\% $\textrm{X/X}_0$. For particle identification, the spectrometer includes a Low‑Gain Avalanche Detector (LGAD)-based Time‑of‑Flight (TOF) system~\cite{Liukang-NST}, while an electromagnetic calorimeter (ECal)~\cite{ECal-NST} is used for neutral‑particle detection. The baseline design of the H‑NS detector is described in detail in Ref.~\cite{Lin:2025yzt}. In this configuration, a carbon foil layer, with a thickness of 0.5 - 1 mm, is placed between tracking detector modules, serving as a scattering target for nucleon polarimetry. As has been demonstrated in BESIII with the carbon fiber in the wall of drift chamber~\cite{PhysRevD.103.092005}, this existing carbon foil can also serve as the $\gamma$-conversion material, with one advantage of not introducing extra material. 

The illustration of $\gamma$-conversion signal at H-NS is displayed in the right plot in Fig.~\ref{fig:HNS_det}. Here the carbon layer severs as the scattering target for the nucleon polarimetry and also photon polarimetry. This study uses a 0.5 mm thick carbon layer as a starting point. Subsequently, layers of different thicknesses are analyzed to determine the optimal value for photon polarization measurement. Three silicon layers after the carbon foil are the minimum requirement to reconstruct the electron and positron of the pair production. To determine the azimuthal angles, the direction of the photon momentum must be known. In this study, we focus on photons originating from the primary vertex. With a known origin, the photon’s flight direction can then be reconstructed from the resulting particle pair, which also enables precise determination of both the conversion vertex and the photon energy.

\begin{figure}
\includegraphics[width=0.4\textwidth]{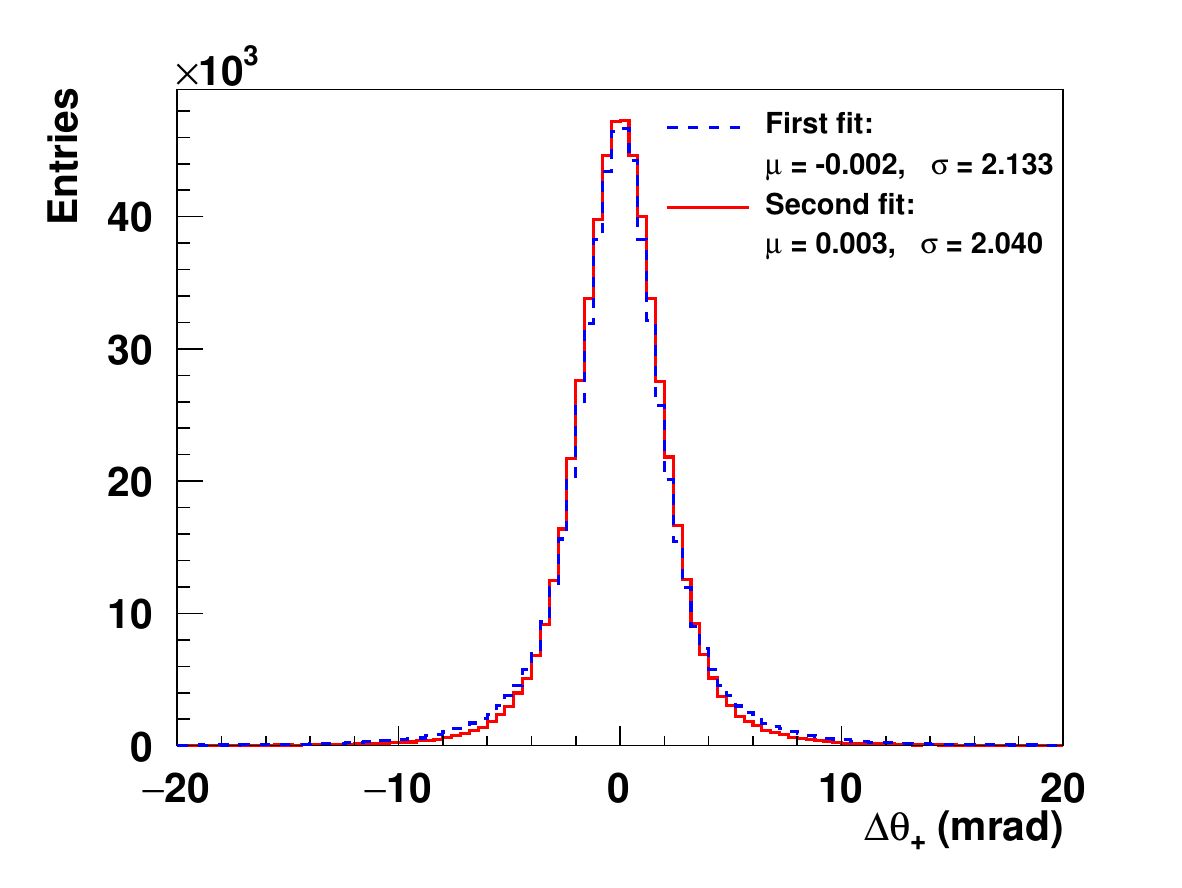}
\includegraphics[width=0.4\textwidth]{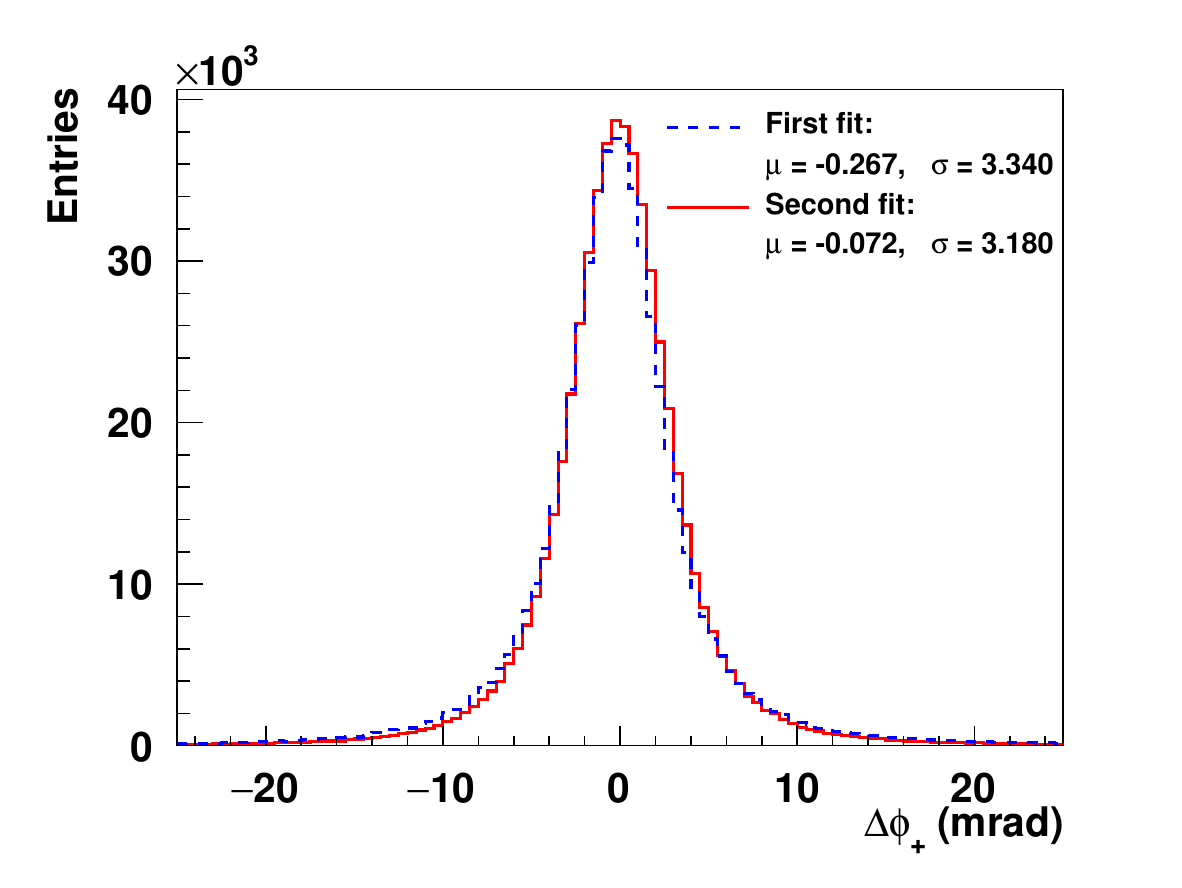}
\caption{\label{fig:H-NS_theta_res} {Angular momentum resolution $\Delta\theta$ (Left) and $\Delta\phi$ (Right) for positron from photon energy of 1 GeV. The dashed blue line represents the first fit using three silicon layers after the carbon foil. The solid red line shows the second iteration with a common vertex with electron track applied. 
}}
\end{figure}


A realistic investigation is carried out using full {\sc Geant4}~\cite{AGOSTINELLI2003250} simulations. In the first stage of the study, pair‑production events are generated with a polarized Bethe–Heitler event generator for
$\gamma$-ray conversion to $e^+e^-$ pairs. This generator employs five‑dimensional probability density functions~\cite{BERNARD201885}. The $e^+e^-$ pairs produced via $\gamma$-conversion are propagated through the H-NS detector using the {\sc Geant4} package ~\cite{AGOSTINELLI2003250} within the HnsRoot software framework based on FairRoot~\cite{Al-Turany:2012zfk}. The output from the {\sc Geant4} simulation—a list of pixel hits—is processed through a track‑fitting procedure implemented with the GenFit package~\cite{HOPPNER2010518}. 



Figure~\ref{fig:H-NS_theta_res} shows the angular resolution for positrons in the $\theta$ and $\phi$ directions from a pair production of photon energy of 1 GeV. The angular resolution of about 2-3 mrad is achieved. Events of $\gamma$‑conversion exhibit a distinct signature: the $e^+$ and $e^-$ originate from a common vertex in the carbon foil. At H-NS, the distance between the reconstructed vertex and the MC truth is extracted and a peak around 80 $\mu$m is observed. Since the angular resolution of the $e^+$ and $e^-$ is crucial for measuring photon polarization, a second iteration of the track fitting—incorporating a common‑vertex constraint—is applied. This constrained fit improves the angular resolution, as shown by the red histograms in Fig.~\ref{fig:H-NS_theta_res}.


\begin{figure}
\includegraphics[width=0.48\textwidth]{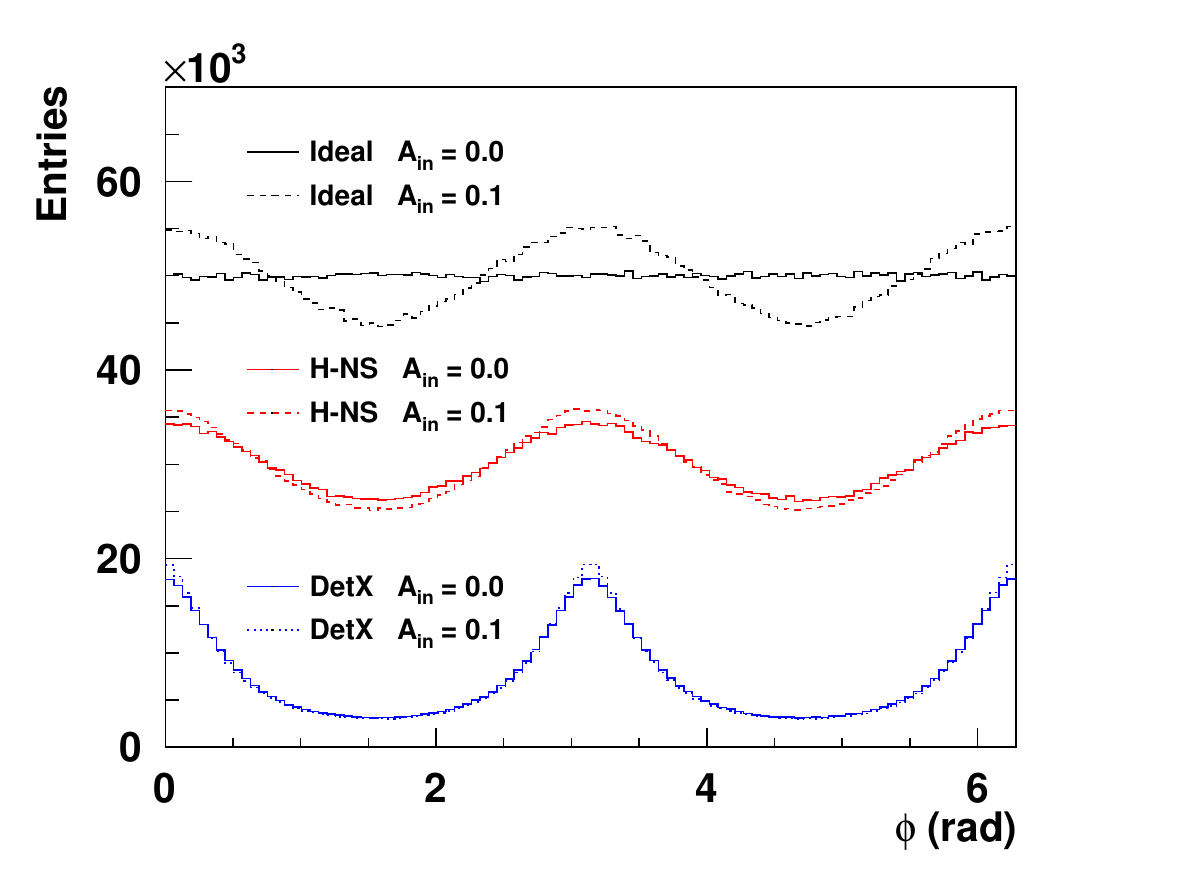}
\caption{\label{fig:phi_6_config} {The smear effect on azimuthal distributions under different detector configurations, including an ideal detector (black lines), H-NS (red lines) and another detector configuration (DetX). For clarity in comparison, the H-NS and DetX distributions have been scaled.
}}
\end{figure}


To measure the photon polarization, one can either fit the azimuthal angle distribution shown in Fig.~\ref{fig:smear_effect}, or apply a moments method with $\mathcal{A}P = 2\sqrt{\langle\cos2\phi\rangle^2 + \langle\sin2\phi\rangle^2}$ and $\phi_0 = \frac{1}{2}\arctan\left(\frac{\langle\sin2\phi\rangle}{\langle\cos2\phi\rangle}\right)$~\cite{Gros:2016zst,Bernard:2013jea}. It should be noted that general-purpose spectrometers with solenoid magnets possess cylindrical symmetry, which can be described in terms of angular resolutions in the $\theta$ and $\phi$ directions. The pair production from photon conversion on a conical surface, combined with detector effects, naturally distorts the azimuthal distribution. To illustrate this, Fig.~\ref{fig:phi_6_config} displays the smear effect on azimuthal distributions under different detector configurations. As a baseline, an ideal detector with perfect angular resolution ($\Delta\theta = 0$ mrad, $\Delta\phi = 0$ mrad) is initially used. The black solid line shows the azimuthal distribution for an ideal detector with an input asymmetry of zero, yielding the expected uniform distribution. In a realistic detector such as H-NS ($\Delta\theta = 2$ mrad and $\Delta\phi = 3$ mrad) the azimuthal distribution at zero input asymmetry is distorted and deviates significantly from a flat line, as indicated by the red solid line. The resulting distorted shape can resemble the functional form of a polarization signal, $1 + \mathcal{A} P \cos 2\phi$. Under different detector configurations, particularly when $\Delta\theta$ differs substantially from $\Delta\phi$, the distortion caused by detector effects becomes more pronounced. This is illustrated by the blue solid line (labeled DetX) in Fig.~\ref{fig:phi_6_config}, where $\Delta\theta = 0$ mrad and $\Delta\phi = 2$ mrad. Nevertheless, by varying the input polarization, detectable differences in the output asymmetry still emerge, as shown by the dashed lines with input asymmetry of 0.1 in Fig.~\ref{fig:phi_6_config}. This confirms that even a non-ideal detector retains sensitivity to photon polarization.

\begin{figure}
\includegraphics[width=0.48\textwidth]{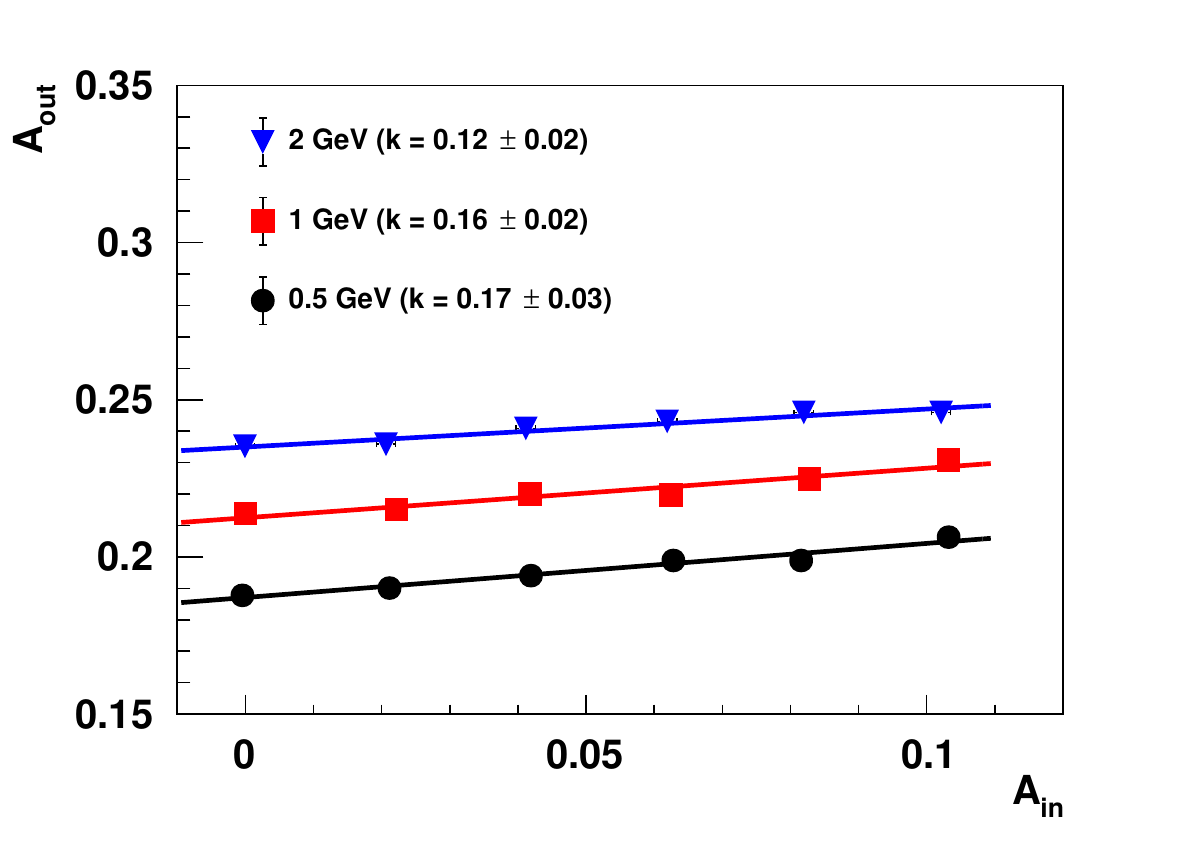}
\caption{\label{fig:Asy_diff_photon_energy} {The extracted asymmetry versus the input asymmetry for different photon energy at H-NS. The red squares are with photon energy of 1 GeV. The black circles and blue triangles are for photon energies of 0.5 GeV and 2 GeV, respectively. A linear fit is applied, with the resulting slope (k) indicated in the brackets.
}}
\end{figure}

To demonstrate the capability of photon polarization measurements, a series of full simulations were performed. In each simulation, one million signal events were generated for different photon energies and different input photon polarization using the $\gamma$-conversion generator~\cite{BERNARD201885}. As a case study, we focus on the barrel part of H-NS which has better angular resolution. MC events are generated within the polar angle range of 30$^\circ$ to 150$^\circ$. The polarization asymmetry of the process, $\mathcal{A}$, is embedded and calculated within the generator itself. As an illustration of the method, we focus on the relationship between the input asymmetry and the extracted asymmetry. For each simulation, the input asymmetry was obtained from a fit to the MC truth of the data sample. The results are displayed in Fig.~\ref{fig:Asy_diff_photon_energy}, where the red squares represent the extracted asymmetries after full detector simulation at H-NS versus the input asymmetries for a photon energy of 1 GeV. A linear fit describes the data well, although the slope is significantly shallower than unity and a nonzero intercept appears. This relationship between the extracted and input asymmetry can therefore serve as a calibration to determine the true asymmetry.

It is important to note that this calibration is dependent on the photon’s energy and its direction of flight. Photons of different energies produce $e^+e^-$ pairs with different opening angles and momenta, leading to variations in the angular resolution of the reconstructed photon momentum. Additionally, photons detected at different polar angles within the laboratory frame exhibit slightly different angular resolutions. As an example, taking the photon energy dependence, the black circles and blue triangles in Fig.~\ref{fig:Asy_diff_photon_energy} show the calibration results for photon energies of 0.5 GeV and 2 GeV, respectively.


\section{Key factors to optimize}
\begin{figure}
\includegraphics[width=0.9\textwidth]{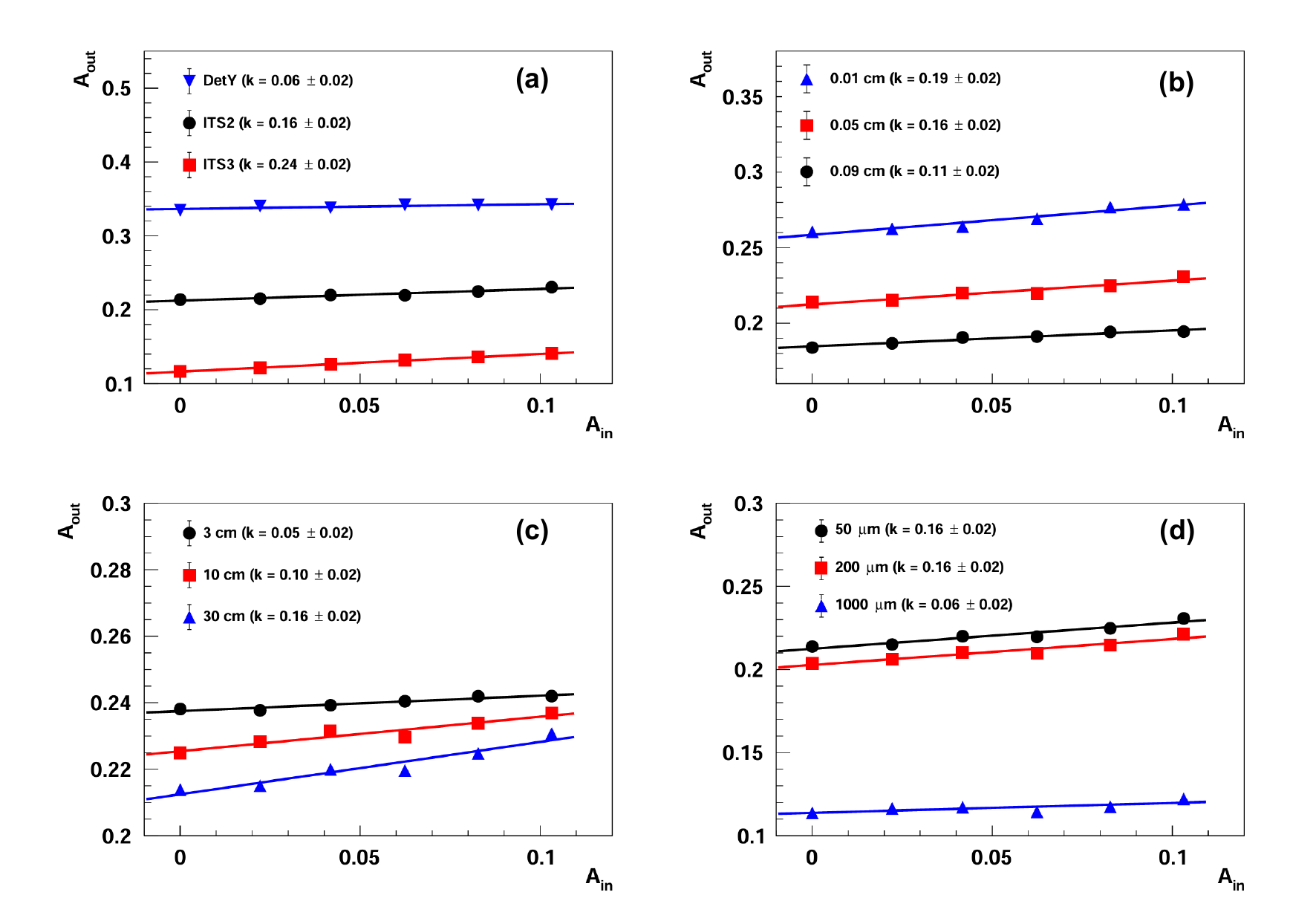}
\caption{\label{fig:factor_optimization} {Key factors affect the capability of photon polarization measurements. Plot (a) represents the extracted asymmetry versus the input asymmetry for different detector technology. Plot (b) shows that with different carbon thickness. Plot (c) is with different carbon location. Plot (d) shows the impact of different primary vertex resolution.
}}
\end{figure}

With the method established, several key factors can be optimized to enhance polarization measurement capability or to better understand their influence. 
As noted, the detector’s angular resolution is critical for photon polarization measurements. The current H-NS design employs three layers of silicon pixel detectors after the carbon foil, each with a pixel size of 30 $\mu$m and a material budget of 0.35\% $\textrm{X/X}_0$ per layer. This is based on mature ALICE ITS2 technology~\cite{Isakov:2025dqm}, successfully used in the ALICE experiment. An upgraded version, ITS3~\cite{LIU2024169355}, will further improve granularity and reduce material budget using a bent-stitched technique, achieving a pixel size down to 20 $\mu$m (or smaller) and an extremely low material budget of 0.05\% $\textrm{X/X}_0$ per layer. 
Figure~\ref{fig:factor_optimization} (a) presents calibration results for different detector technologies in H-NS. The black circles correspond to the nominal H-NS design. The red squares show results with the improved ITS3 technology, exhibiting a steeper slope as expected. For comparison, a third detector configuration—marked as DetY—was tested to represent a worse-case scenario, with a pixel size of 150 $\mu$m and a material budget of 1.5\% $\textrm{X/X}_0$. In this configuration, the angular resolution degrades to approximately 6 mrad in $\phi$ and 3.4 mrad in $\theta$. The corresponding calibration, indicated by the blue triangles in Fig.~\ref{fig:factor_optimization} (a), shows a significantly shallower slope.


\begin{figure}
\includegraphics[width=0.48\textwidth]{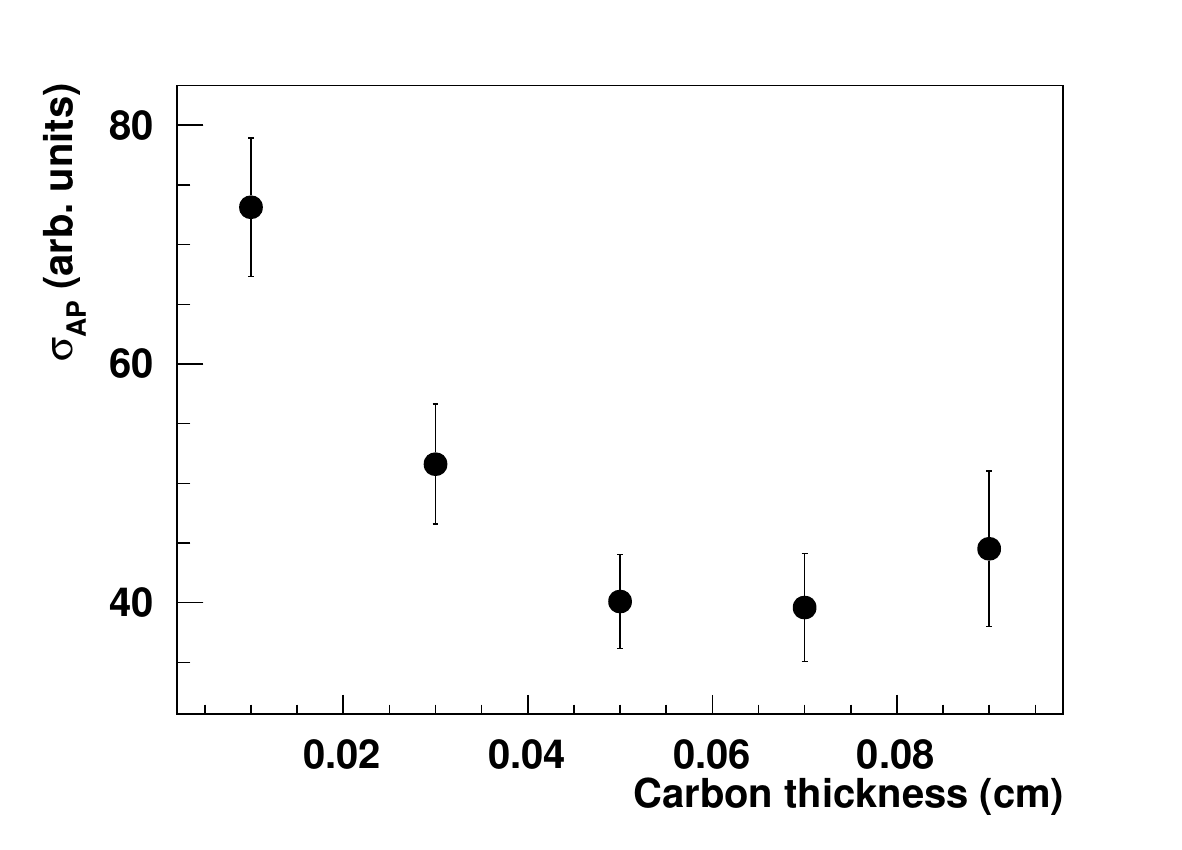}
\caption{\label{fig:AsyErr_diff_carbon_thickness} {Uncertainty estimation at different carbon thickness in arbitrary units.
}}
\end{figure}

In addition to the direct impact of angular resolution, other factors influence photon polarization measurements. These include the thickness of the converter, its location within the experimental setup, and the effects of imprecise primary vertex reconstruction. In the current H-NS nucleon polarimetry design, the use of a carbon foil with a thickness of 0.5–1 mm has been proposed. Figure~\ref{fig:factor_optimization} (b) presents the extracted asymmetry versus the input asymmetry for different carbon thicknesses. A thinner foil can increase the slope in the calibration table because of better angular resolution, but also reduce the conversion probability, requiring a balance between these factors.
As will be detailed in the uncertainty estimation section, the statistical uncertainty of the asymmetry follows $\sigma \approx \sqrt{\frac{2}{N}}/f_{\textbf{slope}} \propto  \sqrt{\frac{2}{d}}/f_{\textbf{slope}}$. Here, $d$ is the foil thickness. An optimization based on this uncertainty criterion has been performed. Figure~\ref{fig:AsyErr_diff_carbon_thickness} plots $\sigma_{\mathcal{A}P} \approx \sqrt{\frac{2}{d}}/f_{\textbf{slope}}$ (in arbitrary units) for different carbon thicknesses. The minimum uncertainty was found to occur at approximately 0.5–0.7 mm.  This indicates that thicknesses in this range are favorable for photon polarimetry and align with the current H-NS design.



In the baseline H-NS design, the carbon foil is positioned at a radius of 30 cm within the barrel. Given the high precision of the $\gamma$-conversion vertex resolution (80 $\mu$m), this distance is sufficient to ensure an accurate determination of the photon direction. A straightforward calculation shows that the resulting photon angular resolution from the vertex uncertainty is on the order of 0.3 mrad (80 $\mu$m / 30 cm), which remains negligible compared to the opening angle from pair production and the angular resolution of the electron and positron momentum reconstruction. We conducted a study by relocating the carbon foil to radii of 3 cm and 10 cm. As shown in Fig.~\ref{fig:factor_optimization} (c), the calibration slope at a distance of 3 cm is much smaller than that at 30 cm, as expected. This indicates that the carbon foil with a larger distance to the primary vertex is favorable.





The determination of photon direction also relies on a reference starting point. For photons originating from the primary vertex, an imprecise vertex position directly affects the reconstructed direction. Modern experiments typically achieve a primary vertex resolution better than 1 mm. Those designed or equipped with dedicated vertex detectors commonly reach a resolution on the order of a few tens of micrometers~\cite{Anderle:2021wcy, AbdulKhalek:2021gbh, KROHN2020164269}. In this study, a primary vertex resolution of 50 $\mu$m is assumed. To evaluate its impact, an additional study was performed by varying the primary vertex resolution. As shown in Fig.~\ref{fig:factor_optimization} (d), the calibration results are compared for primary vertex resolutions of 50 $\mu$m, 200 $\mu$m, and 1000 $\mu$m. The plot indicates that the calibration remains stable for resolutions up to 200 $\mu$m, while a degraded resolution of 1000 $\mu$m reduces the calibration slope significantly.





\section{Probability of $\gamma$-conversion and uncertainty estimation}


For carbon foil of 0.5 mm thickness, the corresponding material budget is about 0.25\% $\textrm{X/X}_0$. It's impact on the conventional detector performance has been found negligible as detailed in Ref~\cite{Lin:2025yzt}. The probability of $\gamma$-conversion to $e^+e^-$ pair in the carbon foil is around 0.002. In the photon polarization measurement, the asymmetry $\mathcal{A}P$ is extracted from the azimuthal distribution, and the statistical uncertainty of the observed $\mathcal{A}P$ is expressed as ~\cite{BERNARD2013765}



\begin{align}   
\sigma_{\mathcal{A}P}^{\textbf{obs}} \approx \sqrt{\frac{2-(\mathcal{A}P)^2}{N}} \approx \sqrt{\frac{2}{N}},
\end{align}

Here, $N$ is the number of observed signal events from $\gamma$-conversion. The expression on the right side of the equation is a good approximation due to the small value of $\mathcal{A}P$~\cite{WOJTSEKHOWSKI2003605}. The true asymmetry $\mathcal{A}P$ is obtained by dividing the fitted slope in the corresponding calibration table. 
Consequently, the uncertainty in $\mathcal{A}P$ is amplified by a factor of 1/$f_{\textbf{slope}}$. For detectors with poor angular resolution, this factor can become very large, introducing a significant uncertainty that is undesirable.

\begin{align}   
\sigma_{\mathcal{A}P}^{} =  \sigma_{\mathcal{A}P}^{\textbf{obs}} / f_{\textbf{slope}} \approx \sqrt{\frac{2}{N}} / f_{\textbf{slope}} ,
\end{align}


Statistically, it is possible to measure photon polarization in high-luminosity collider experiments. Taking the $e^+e^-$ collider as an example, the current BESIII experiment has accumulated 10 billion $J/\psi$ events and 3 billion $\psi(3686)$ events, and the future STCF~\cite{Achasov:2023gey} is expected to increase the luminosity by a factor of 100. Many decay channels from these charmonium states involve photons. For a channel with a branching fraction of $10^{-3}$, assuming a $\gamma$-conversion rate of 0.002 (for a carbon foil of 0.5 mm thickness) and a detector efficiency of 50\%, the number of $\gamma$-conversion signal events is roughly $1\times10^4$ at BESIII and $1\times10^6$ at STCF. This yields an expected statistical uncertainty on the observed asymmetry of about 1.4\% for BESIII and 0.14\% for STCF. STCF is planned to be equipped with a vertex detector. Assuming it can achieve an angular resolution of 2-3 mrad in both the $\theta$ and $\phi$ directions --- comparable to that of the H-NS --- it would have a calibration slope of approximately 0.15. As a result, the final uncertainty on the true asymmetry would be on the order of 1.0\%, which is sufficiently precise to perform reliable measurements. 

Throughout the paper, we focus on the observed asymmetry, $\mathcal{A}P$. Determining the photon's linear polarization $P$ requires prior knowledge of the process's polarization asymmetry $\mathcal{A}$, which can be derived from the differential cross-section equations. In the GeV energy range, $\mathcal{A}$ typically lies between 0.1 and 0.2 and depends on the energy fraction carried by the electron or positron relative to the photon. The small magnitude of $\mathcal{A}$ will amplify the uncertainties in the final extracted photon polarization.


Beyond statistical uncertainty, the process of estimating systematic uncertainty in photon polarization measurements is complex. Given a fixed spectrometer design, the angular resolution (in $\theta$ and $\phi$) can be determined with a certain uncertainty. This uncertainty in angular resolution introduces systematic uncertainty into the calibration table. Moreover, the calibration table itself depends on photon energy, flight direction, and the opening angle distribution of the $e^+e^-$ pair. These dependencies propagate their own uncertainties into the calibration. All such systematic uncertainties must be accounted for in future experiments.





    

\section{Summary}

In the paper, we propose a method for measuring photon polarization using a general-purpose spectrometer without compromising its conventional performance. The method is validated through realistic {\sc Geant4} simulations based on the H-NS detector,
demonstrating that modern general-purpose spectrometers can feasibly function as photon polarimeters. The core technique involves compiling a calibration table to map the observed polarization values to their corresponding true values. This calibration table depends on photon energy, the spectrometer's angular resolution, the resolution of primary vertex and the location and thickness of the scattering layer. A poorer angular resolution results in an amplified uncertainty and requiring more statistics to compensate. Therefore, improved angular resolution is highly beneficial for polarization measurements. With current technology, silicon pixel detectors with a pitch size of 20-30 $\mu$m are now mature and widely used in many experiments. Employing a silicon detector, could enhance angular resolution and improve performance in the photon polarimetry role. 


Although H-NS is a fixed-target experiment designed for collisions of $pp$, $pA$, or $AA$, its spectrometer follows a general design that is also applicable to collider experiments, including those that involve scattering of $e^+e^-$ and $ep$. In this paper we present detailed optimizations of the photon polarimetry capability. Beyond the strong dependence on angular resolution, the study also identifies key factors affecting polarization measurements, such as the optimization procedure of location and thickness of the scattering layer and the influence of primary vertex resolution. The method is validated with a solenoid based spectrometer, while it is in principle also applicable on dipole based spectrometers, such as CBM~\cite{KLOCHKOV2021121945}, etc. These findings establish a valuable benchmark for existing and future experiments, such as BESIII~\cite{ABLIKIM2010345}, Belle-II~\cite{ONUKI202278}, STCF~\cite{Achasov:2023gey}, EicC~\cite{Anderle:2021wcy}, EIC~\cite{AbdulKhalek:2021gbh}, etc. The proposed method enables the measurement of final-state photon polarization with general-purpose spectrometers. This additional polarization information can offer new insights into nuclear and particle physics.


\begin{acknowledgments}
This work is supported in part by the National Key Research and Development Program of China under Contract No. 2023YFA1606800, 2024YFA1611000, the National Natural Science Foundation of China (NSFC) under Contract No. 11975278.
 
\end{acknowledgments}

\bibliography{polarimeter}

\end{document}